\documentclass[prd,aps,preprint]{revtex4}

\usepackage{amssymb}
\usepackage{epsfig}
\usepackage{subfigure}
\usepackage{color}
\usepackage{array}
\usepackage{url}
\input epsf.tex
\def\DESepsf(#1 width #2){\epsfxsize=#2 \epsfbox{#1}}

\begin{document}


\preprint{\vbox{ \hbox{MIFP-09-45, UMD-PP-09-059} }}

\title{{\Large\bf  An SO(10) Grand Unified Theory of Flavor}}
\author{{\bf Bhaskar Dutta}$^1$, {\bf Yukihiro Mimura}$^2$ and {\bf R.N.
Mohapatra}$^2$ }

\affiliation{
$^1$Department of Physics, Texas A\&M University,
College Station, TX 77843-4242, USA
\\
$^2$ Maryland Center for Fundamental Physics and Department of Physics,
University of Maryland, College Park, MD, 20742, USA}
\date{November, 2009}

\begin{abstract}
We present a supersymmetric SO(10) grand unified theory (GUT) of
flavor based on an $S_4$ family symmetry. It makes use of our
recent proposal to use SO(10) with type II seesaw mechanism for
neutrino masses combined with a simple ansatz that the dominant
Yukawa matrix (the {\bf 10}-Higgs coupling to matter) has rank
one. In this paper, we show how the rank one model can arise
within some plausible assumptions as an effective field theory
from vectorlike {\bf 16} dimensional matter fields with masses
above the GUT scale. In order to obtain the desired fermion flavor
texture we use $S_4$ flavon multiplets which acquire vevs in the
ground state of the theory. By supplementing the $S_4$ theory with
an additional discrete symmetry, we find that the flavon  vacuum
field alignments take a discrete set of values provided some of
the higher dimensional couplings are small. Choosing a particular
set of these vacuum alignments appears to lead to an unified
understanding of observed quark-lepton flavor:
 (i) the lepton
mixing matrix that is dominantly tri-bi-maximal with small corrections
related to quark mixings; (ii) quark lepton mass relations at GUT
scale: $m_b\simeq m_{\tau}$ and $m_\mu\simeq 3 m_s$ and (iii) the
solar to atmospheric neutrino mass ratio $m_\odot/m_{\rm atm}\simeq
\theta_{\rm Cabibbo}$ in agreement with observations. The model
predicts the neutrino mixing parameter, $U_{e3}
\simeq \theta_{\rm Cabibbo}/(3\sqrt2) \sim 0.05$, which should be observable in planned
long baseline experiments.
 \end{abstract}


\maketitle

\vskip1.0in

\newpage

\baselineskip 18pt

\section{Introduction}
A unified understanding of the diverse pattern of quark lepton
masses and mixings is a fundamental challenge for physics beyond
the standard model~\cite{weinberg}. The two major elements of this
flavor puzzle that any theory must explain are: (i) strong mass
hierarchy in the quark and charged lepton sector and weak
hierarchy for neutrinos; (ii) large lepton mixings i.e.
$\theta^l_{23}\sim 45^{\rm o}$ and $\theta^l_{12}\simeq 35^{\rm
o}$ as against small quark mixings $\theta^q_{23}\sim 2.5^{\rm o}$
and $\theta_{12}^q\sim 13^{\rm o}$ and apparent relation between
some of the mixing angles and the fermion masses. Since grand
unified theories (GUT) not only unify different gauge couplings at
a high scale  but also unify quarks and leptons within a single
framework, they have often been thought of as an attractive venue
for unraveling this puzzle. Furthermore the fact that the seesaw
mechanism for understanding small neutrino masses~\cite{seesaw}
also seems to require a $B-L$ breaking scale close to the scale of
coupling unification, makes this suggestion quite promising. The
constraints of higher symmetry however make it highly nontrivial
to understand all the details of flavor puzzle although many
attempts have been made~\cite{many}.

In a recent paper, we have suggested a possible way~\cite{dmm} to
address this problem in supersymmetric SO(10) GUT models. The main
assumptions of ref.~\cite{dmm} are: (a) all fermion masses arise
from effective Yukawa couplings~\cite{babu} involving {\bf 10} and
{\bf 126} Higgs multiplets; (b) neutrino masses arise~\cite{goran}
from type II seesaw mechanism~\cite{type2} and (c) the {\bf
10}-Higgs Yukawa dominates fermion masses and has rank one. We
showed in ref.~\cite{dmm} how this program when implemented using
the already mentioned Higgs content of a single {\bf 10}, {\bf
126} plus possibly another {\bf 10} or {\bf 120} Higgs fields  not
only explains all the qualitative features of quark and lepton
flavor noted above but also makes a prediction for the lepton
mixing angle $U_{e3}$ or $\theta_{13}$. In most models we
discussed in \cite{dmm}, the apparent tri-bi-maximal mixing
pattern~\cite{tbm} observed for neutrinos did not arise from any
symmetry. In this note, we pursue program outlined in \cite{dmm}
further by using discrete family symmetries to make this ansatz
more predictive. Our strategy is to use flavon fields whose vevs
give the effective Yukawa couplings responsible for fermion masses
at the GUT scale. We use additional discrete family symmetries
whose role is to constrain the ground state of the flavon
Hamiltonian such that they lead to particular textures for the
fermion mass matrices within certain assumptions. We are able to
isolate a set of allowed flavon vacuum states which are such that
the dominant part of the lepton mixing matrix naturally has a
tri-bi-maximal form, provided some of the higher dimensional terms
in the flavon superpotential are small. The desired flavon vacuum
alignment seems to arise naturally with an $S_4$
symmetry~\cite{hagedorn} which unifies all three families of
fermions into a ${\bf 3_2}$ multiplet.

The new results of this paper are : (i) we show how the rank one
model can arise naturally  as an effective field theory from
vectorlike {\bf 16} dimensional matter fields with masses above
the GUT scale; and (ii) how the detailed fermion flavor textures
arise from the vacuum field alignments  of gauge singlet $S_4$
flavon fields leading to the following results naturally without
adjustment of parameters: (a) the lepton mixing matrix has
dominantly tri-bi-maximal form  with  small corrections related to
quark mixings; (b) quark lepton mass relations at GUT scale:
$m_b\sim m_{\tau}$ and $m_\mu\simeq 3 m_s$ and (c) the solar to
atmospheric mass ratio $m_\odot/m_{\rm atm}\simeq \theta_{\rm
Cabibbo}$ in agreement with observations.

\section{Overview of the SUSY SO(10) rank one strategy}
We use the Higgs fields that give fermion masses to consist of two
{\bf 10} dimensional multiplets (denoted by $H, H'$) and a single
${\bf 126}+\overline{\bf 126}$ (denoted by $\Delta$ and
$\overline{\Delta}$). The Yukawa superpotential for this case in a
generic SO(10) model can be written as:
\begin{eqnarray}
W_Y~ =~h\, \psi\psi H + f\, \psi\psi\bar{\Delta}+h'\,\psi\psi \, H'\,,
\end{eqnarray}
where the symbol $\psi$ stands for the {\bf 16} dimensional
representation of SO(10) that represents the matter fields.
The coupling matrices $h, h'$ and $f$ are symmetric. As we show
later in this paper, their detailed texture will be determined by
the $S_4$ symmetry.
The representations $H$, $H'$ and $\Delta$ each have two standard
model (SM) doublets in them. The general way to understand
how the two MSSM doublets arise from them is as follows: at the
GUT scale $M_U$, after the GUT and the $B-L$ symmetries are
broken, one linear combination of the up-type doublets and one of
down-type ones remain almost massless whereas the remaining ones
acquire GUT scale masses just like the color triplet and other
non-MSSM multiplets. The electroweak symmetry is broken after the
light MSSM doublets (to be called $H_{u,d}$) acquire vacuum
expectation values (vevs) and they then generate the fermion
masses. The resulting formulae for different fermion masses are
given by:
\begin{eqnarray}
Y_u &=& h + r_2 f +r_3 h^\prime, \label{eq2} \\\nonumber
Y_d &=& r_1 (h+ f + h^\prime)\,, \\\nonumber
Y_e &=& r_1 (h-3f + c_e h^\prime)\,, \\\nonumber
Y_{\nu^D} &=& h-3 r_2 f + c_\nu h^\prime,
\end{eqnarray}
where $Y_a$ are mass matrices divided by the electro-weak vev
$v_{wk}$ and $r_i$ and $c_{e,\nu}$ are the mixing parameters which
relate the $H_{u,d}$ to the doublets in the various GUT
multiplets.
More precisely, the matrices $h$, $f$ and $h^\prime$ in $Y_a$ are
multiplied by the Higgs mixing parameters when they appear in the fermion
mass matrices. The definitions of the couplings and the Higgs
mixing parameters are given in ref.~\cite{mimura}. In our particular case
with a second {\bf 10}-Higgs ($H^\prime$), $c_e= 1$ and $c_\nu =
r_3$.
Furthermore, we use the type II seesaw formula for getting
neutrino masses which is possible to obtain with symmetry breaking
pattern in SO(10) as given in~\cite{goh}.
\begin{eqnarray}
{\cal M}_\nu~=~fv_L.
\label{eq4}
\end{eqnarray}
Note that $f$ is the same coupling matrix that appears in the
charged fermion masses in Eq. (\ref{eq2}), up to factors from the
Higgs mixings and the Clebsch-Gordan coefficients. This helps us to
 connect the neutrino parameters to the quark-sector parameters. The equations
(\ref{eq2}) and (\ref{eq4}) are the key equations in our unified
approach to addressing the flavor problem.

The main hypothesis of our approach in ref.~\cite{dmm} is that
\begin{itemize}
\item  the fermion mass formula of Eq. (\ref{eq2}) are dominated
by the matrix $h$ with the contributions of $f$ and $h'$ being
small perturbations;

\item  the matrix $h$ has rank one.

\end{itemize}

It follows from these assumptions that in the limit of $f,h'\to
0$, the quark and lepton mixings vanish as do the neutrino masses.
Once $f,h'$ are turned on, one can choose $f$ to be diagonal by an
appropriate choice of basis and without any loss of generality.
Since the neutrino masses are diagonal in this basis, the entire
Pontecorvo-Maki-Nakagawa-Sakata (PMNS) matrix comes from the matrix
that diagonalizes the charged
lepton mass matrix and for arbitrary form of the later, the PMNS
matrix will in general have large mixing angles.
On the other hand, the Cabibbo-Kobayashi-Maskawa (CKM) matrix
$V_{\rm CKM}=U^\dagger_u U_d$
which in the limit of $f,h'\to 0$ is equal to a unit matrix, owes the
origin of quark mixings to $f,h'$.
%
%
The quark mixings are
proportional to $|f|/|h|$ and hence small as observed. It is also
clear that the charged lepton and quark masses of second and first
generation are also proportional to $|f|/|h|$ and thus
hierarchical.

Our procedure in this paper is as follows: we supplement the above
rank one hypothesis by a discrete family symmetry $S_4$ so that
forms of the $h, f, h'$ are consequences of the vacuum expectation
values of gauge singlet but $S_4$ non-singlet flavon fields,
$\phi_i$, thereby making the model more predictive. To implement
this procedure, we first derive the GUT scale
 effective Lagrangian from a
pre-GUT scale theory that has vectorlike {\bf 16}-dim. matter
spinor with masses slightly above the GUT scale.  The resulting effective
theory involves non-renormalizable higher dimensional operators involving
$\psi$, Higgs fields and the flavon fields $\phi_i$ whose vevs generate
the flavor texture observed at GUT scale. These are then extrapolated to
the weak scale to compare with observations.

\section{$S_4$ family symmetry and model of flavor}
The $S_4$ group is a 24 element group describing permutations of four
distinct objects and has five irreducible representations with
dimensions ${\bf 3_1\oplus 3_2 \oplus 2 \oplus 1_2 \oplus 1_1}$.
The distinction between the representations with subscripts $1$
and $2$ is that the later change sign under the transformation of
group elements involving the odd number of permutations of $S_4$.
For other details of $S_4$ group, see~\cite{hagedorn}.

We assign the three families of {\bf 16}-dim. matter fermions $\psi$ to
${\bf 3_2}$-dim. representation of $S_4$ and the Higgs field $H$,
$\bar{\Delta}$ and $H'$ to 
${\bf 1_1, 1_2}$, and ${\bf1_1}$ reps, respectively.
We then choose three 
SO(10) singlet flavons $\phi_i$ transforming
as ${\bf 3_2, 3_1, 3_2}$ reps of $S_4$ and
one gauge and $S_4$ singlet fields $s_1, s_2$ transforming as $\bf 1_2$ and $\bf 1_1$
 respectively.
%
%
We further assume that at
a scale slightly above the GUT scale, there are two $S_4$ singlet
vectorlike pairs of ${\bf 16\oplus \overline{16}}$ fields denoted by
$\psi_{V}$ and $\bar{\psi}_V$. In order to get the desired Yukawa
couplings naturally from this high scale theory, we supplement the
$S_4$ group by an $Z_n$ 
group with all the above fields
belonging to representations given in the Table I.
\begin{table}
\begin{tabular}{|c|c|c|c|c|c|c|c|c|c|c|c|c|c|c|c|c|
} \hline
& $\psi$ & $H$ & $\bar\Delta$ & $H^\prime$ & $\phi_1$ & $\bar\phi_1$ & $\phi_2$ &
$\bar\phi_2$ & $\phi_3$ & $\bar\phi_3$ & $\psi_{V1}$ & $\bar\psi_{V1}$ & $\psi_{V2}$ &
$\bar\psi_{V2}$ & $s_1$ & $s_2$
\\ \hline
SO(10) & $\bf 16$ & $\bf 10$ & $\overline{\bf 126}$ & $\bf 10$ &
$\bf 1$ & $\bf 1$ & $\bf 1$ & $\bf 1$ & $\bf 1$ & $\bf 1$ & $\bf 16$ & $\overline{\bf 16}$ &
$\bf 16$ & $\overline{\bf 16}$ & $\bf1$ & $\bf1$
\\ \hline
$S_4$ & $\bf 3_2$ & $\bf
1_1$ & $\bf 1_2$ & $\bf 1_1$ & $\bf 3_2$ & $\bf 3_2$ & $\bf 3_1$ & $\bf 3_1$ & $\bf 3_2$ &
$\bf 3_2$ & $\bf 1_1$ & $\bf 1_1$ & $\bf 1_2$ & $\bf 1_2$  & $\bf 1_2$ & $\bf 1_1$
\\ \hline
%
%
$Z_n$ & $1$ & $\omega^{-4}$ & $\omega^{-2-a}$ & $\omega^{-1}$ &
$\omega^{2}$ & $\omega^{-2}$ & $\omega$ & $\omega^{-1}$ &
$\omega^{2+a-b}$ & $\omega^{-2-a+b}$ &
$\omega^2$ & $\omega^{-2}$ & $\omega$ & $\omega^{-1}$ & $\omega^a$ &$\omega^b$
\\ \hline
\end{tabular}
\caption{The fields and representations to generate the desired Yukawa couplings.
$\omega = e^{i \frac{2\pi}{n}}$.}
\end{table}

The most general high scale  Yukawa superpotential involving
matter fields invariant under this symmetry is given by:
\begin{eqnarray}\label{generalYukawa}
W = (\phi_1 \psi)\bar{\psi}_{V1}~+~ \psi_{V1} \psi_{V1} H ~+~M_1
\bar{\psi}_{V1} \psi_{V1}~\\ \nonumber +(\phi_2
\psi)\bar{\psi}_{V2}~+~ \frac{1}{M_P}s_1 \psi_{V2} \psi_{V2}
\bar\Delta +~M_2\bar{\psi}_{V2} \psi_{V2}\\ \nonumber ~+
\frac{1}{M^2_P}s_2 (\phi_3 \psi \psi) \bar \Delta +
\frac{1}{M_P}(\phi_2 \psi \psi) H^\prime,
\end{eqnarray}
where the brackets stand for the $S_4$ singlet contraction of
flavor index. The singlet field $s_i$ can have large vev as follows:
consider its $Z_n$ charge to be such that the only polynomial
term involving the $s_i$ in the superpotential has the form
$s_i^{k_i}/M^{k_i-3}_{P}$
(in order to describe the essential potential, we ignore a possible
$s_1^{\ell_1} s_2^{\ell_2}$ term).
The dominant part of the potential in
the presence of SUSY breaking has the form:
\begin{eqnarray}
V(s_i)~=~-m^2_{s_i} |s_i|^2+k\frac{s_i^{2k_i-2}}{M^{2k_i-6}_P}+\cdots.
\end{eqnarray}
Minimizing this leads to
$\langle s_i \rangle\sim[m^2_{S_i}M^{2k_i-6}_P]^{\frac{1}{2k_i-4}}$, which is above
GUT scale for larger values of the integer $k_i$ (which in turn is
determined by the $Z_n$ symmetry charge of $s_i$).
One could also have large vevs for $s_1, s_2$ by using anomalous $U(1)$
charges for them using $D$-terms to break the $U(1)$ symmetry.

The effective theory below the scales $M_{1,2}$ and $\langle s_i \rangle$ of the
vector-like pair masses and the $s_i$-vevs respectively is given by:
\begin{eqnarray}
W = (\phi_1 \psi) (\phi_1 \psi) H + (\phi_2 \psi) (\phi_2 \psi)
\bar\Delta + (\phi_3 \psi \psi) \bar \Delta + (\phi_2 \psi \psi)
H^\prime,
\label{phiyukawa}
\end{eqnarray}
where we have omitted the dimensional coupling constants to make
it simple for the purpose of writing. The discrete symmetries
prevent $\phi^2/M^2$ corrections to these terms. So our
predictions based on this effective superpotential do not receive
large corrections.
We note that the non-renormalizable terms in
Eq.(\ref{generalYukawa}) can also be obtained from renormalizable
couplings if we introduce further $S_4$-triplet vectorlike fields.
Here, however we use only $S_4$-singlet vectorlike fields to get
rank 1 contribution to $h$ and $f$ Yukawa couplings and that is
why we need the non-renormalizable terms to be present in
Eq.(\ref{generalYukawa}.
%
%
A few comments are in order regarding the need for the extra $Z_n$
symmetry.

\begin{itemize}

\item The $Z_n$ 
 group provides a selection rule of the flavon
couplings and the charges of various fields under this are chosen
so as to forbid direct renormalizable Yukawa coupling, e.g.,
$(\psi \psi) H$, which can lead to loss of rank one property and
hence the hierarchy of fermion masses.
%
\item The barred flavon fields $\bar\phi_i$ are introduced to
obtain the potential of the flavons necessary for our vacuum
alignment. They do not couple to matter fields.
\item We note that the replacement of $\phi_1$ with $\phi_3$,
$\bar\phi_3$ is forbidden if $a-b \neq 0, -4$, and similarly
unwanted terms can be forbidden when $n$ is a large number.

\item The term $\phi_1 \psi\psi \bar\Delta S_1$ is $Z_n$
invariant, but transforms as $\bf 1_2$ under $S_4$ because $\psi
\psi$ is symmetric due to SO(10) algebra and thus it is not
allowed either.
\item The $S_4$ invariant singlet $s_2$ is introduced to forbid
$\phi_2^2 \bar\phi_3$, $\bar\phi_2^2\phi_3$ terms, which are
unwanted in the flavon superpotential.

\item The $Z_n$ symmetry allows mixed higher dimensional terms of
the form $\phi_i\bar\phi_i \phi_j\bar\phi_j$ terms with $i\neq j$.
We assume that the couplings of these terms are small compared to
other terms so that the alignment shift caused by these terms
compared to that given below is small and does not affect our
result.

\end{itemize}

 The details of the flavon superpotential will be discussed later.

In order to get  fermion masses, we have to find the alignment
\cite{Ross} of the vevs of the flavon fields $\phi_{1,2,3}$. We
show below that the following choice of vevs are among the minima
of the flavon superpotential provided the couplings of mixed terms
between different $\phi_i$'s are small compared to other
couplings:
%
%
%
\begin{equation}
\phi_1 = \left(
\begin{array}{c}
0 \\ 0 \\ 1
\end{array}
\right), \quad \phi_2 = \left(
\begin{array}{c}
0 \\ -1 \\ 1
\end{array}
\right), \quad \phi_3 = \left(
\begin{array}{c}
1 \\ 1 \\ 1
\end{array}
\right). \label{flavon}
\end{equation}
Clearly, there are other vacua for the flavon model that we do
not choose. What is however nontrivial is that the alignments are
along quantized directions. This is a consequence of supersymmetry
combined with discrete symmetries in the theory. Given these vev,
we find from Eq. (\ref{phiyukawa}) that the Yukawa coupling
matrices $h,f,h'$ have the form:
\begin{eqnarray}
h &\propto& \left(
\begin{array}{ccc}
0 & 0 & 0 \\
0 & 0 & 0 \\
0 & 0 & 1
\end{array}
\right),
\\
f &\propto& \left(
\begin{array}{ccc}
0 & 0 & 0 \\
0 & 1 & -1 \\
0 & -1 & 1
\end{array}
\right)
 + \lambda
\left(
\begin{array}{ccc}
0 & 1 & 1 \\
1 & 0 & 1 \\
1 & 1 & 0
\end{array}
\right), \\
h^\prime &\propto& \left(
\begin{array}{ccc}
0 & 1 & -1 \\
1 & 0 & 0 \\
-1 & 0 & 0
\end{array}
\right),
\end{eqnarray}
and the charged fermion mass matrices can then be inferred.
 The neutrino mass matrix in this basis has the form:
\begin{equation}
{\cal M}_\nu~=~\left(\begin{array}{ccc} 0 & c & c\\c & a & c-a \\
c & c-a & a\end{array}\right),
\end{equation}
where $c/a = \lambda \ll 1$.
It is diagonalized by the tri-bi-maximal matrix
\begin{equation}
U_{\rm TB} = \left( \begin{array}{ccc}
                 \sqrt{\frac23} & \sqrt{\frac13} & 0 \\
                 -\sqrt{\frac16}  & \sqrt{\frac13}& -\sqrt{\frac12} \\
                 -\sqrt{\frac16} & \sqrt{\frac13} & \sqrt{\frac12}
               \end{array}
        \right).
\end{equation}
 This is however not the full PMNS matrix which will
receive small corrections from diagonalization of the charged
lepton matrix, which not only make small contributions to the
$\theta_{\rm atm}$ and $\theta_{\odot}$ but also generate a small
$\theta_{13}$.

The neutrino masses are given by $m_{\nu3}~=~2a-c\,$;
$m_{\nu2}~=~2c$ and $m_{\nu1}~=~-c$. To fit observations, we
require $\lambda = c/a \simeq \sqrt{\Delta m_{\odot}^2/\Delta
m_{\rm atm}^2} \sim 0.2$, which fixes the neutrino masses
$m_{\nu3} \simeq 0.05$ eV, $m_{\nu2} \simeq 0.01$ eV, and
$m_{\nu1} \simeq 0.005$ eV. We will see below that $\lambda$ is
also the Cabibbo angle substantiating our claim that neutrino mass
ratio and Cabibbo angle are related.


 For the charged lepton, up and down quark mass matrices, we have:
\begin{eqnarray}
M_\ell ~=~\frac{r_1}{\tan\beta}\left(\begin{array}{ccc} 0 & -3m_1 + \delta &
-3m_1-\delta \\ -3m_1 + \delta & -3m_0 & 3m_0-3m_1 \\
-3m_1-\delta & 3m_0-3m_1 & -3m_0+M\end{array}\right),\\ \nonumber
M_d ~=~\frac{r_1}{\tan\beta}\left(\begin{array}{ccc} 0 & m_1 + \delta &
m_1-\delta \\ m_1 + \delta & m_0 & -m_0+m_1 \\
m_1-\delta & -m_0+m_1 & m_0+M\end{array}\right),\\ \nonumber M_u
~=~\left(\begin{array}{ccc} 0 & r_2m_1 + r_3\delta &
r_2m_1-r_3\delta \\ r_2m_1 + r_3 \delta & r_2m_0 & -r_2m_0+r_3m_1 \\
r_2m_1+r_3\delta & -r_2m_0+r_3m_1 & r_2m_0+M\end{array}\right),
\end{eqnarray}
where $\tan\beta$ is a ratio of $H_{u,d}$ vevs.
Note that $m_1/m_0 = \lambda \sim 0.2$ and of course $m_0 \ll M$.
A quick examination of these mass matrices leads to several
immediate conclusions:

\begin{enumerate}

\item The model predicts that at GUT scale $m_b\simeq m_\tau$.

\item Since $(M_d)_{11} \to 0$, we get $V_{us} \simeq
\sqrt{m_d/m_s}$.

\item The empirically satisfied relation $m_\mu m_e\simeq m_s m_d$
can be obtained by the choice of parameters $ -3m_1+\delta  =
(m_1+\delta)e^{i\sigma}$, where $\sigma$ is a phase. Solving this
equation, we find that $\delta = m_1 (1+i \cot \sigma/2)$.
We obtain $V_{us} \simeq (1-r_3/r_2) \delta/m_0$,
thereby relating Cabibbo angle to the neutrino mass ratio $m_{\odot}/m_{\rm atm}\simeq \lambda$.

\item $m_\mu\sim -3 m_s$.

\item The leptonic mixing angle to diagonalize $M_\ell$
is related to quark mixing $\theta^l_{12}\sim \frac13
 V_{us}$, which leads to a prediction for $\sin\theta_{13}\equiv
 U_{e3}\sim \frac{V_{us}}{3\sqrt{2}}\simeq 0.05$ \cite{king}.

 \item  $\displaystyle V_{cb}\sim \frac{m_s}{m_b} \cot \theta_{\rm atm}$.

\item The masses of up and charm quarks are given by the parameters
$r_{2,3}$ and are therefore not predictions of the model.

\item CP violation in quark sector can put in by making the
parameters $h'$ complex.

\item
The model predicts a small amplitude for neutrino-less double
beta decay from light neutrino mass:
$m_{\nu_{ee}} \sim c \sin\theta_{12}^l \simeq 0.3$ meV.

 \end{enumerate}

 The first four relations are fairly well satisfied by
 observations; the fifth  prediction (i.e. that for $U_{e3}$)
  can be tested in upcoming reactor and
  long baseline experiments. Note that the deviation from
  tri-bi-maximal mixing pattern coming from the charged lepton
  mass diagonalization could be thought of as a small perturbation of
  the neutrino mass matrix \cite{werner} except that we predict
  the form of the perturbation from symmetry considerations.
The sixth prediction gives a smaller value for $V_{cb}$ ($0.02$ as
against observed GUT scale value of $0.03$) if one uses GUT scale
extrapolated value of the known $b$ mass. However, in the MSSM there
are threshold corrections to the $b-s$ quark mass mixing from
gluino and wino exchange one-loop diagrams; by choosing this
contribution, one could obtain the desired $V_{cb}$.

Note that in this model, the top quark Yukawa coupling at GUT scale
 arises from an effective higher dimensional operator.
We have showed the effective operator in Eq.(\ref{phiyukawa}) by
expanding $\phi/M$. The more precise form for the top Yukawa
coupling is $\phi^2/(M_1^2+\phi^2) h_{\psi_V \psi_V H}$, where
$h_{\psi_V \psi_V H}$ is a coupling of $\psi_V \psi_V H$ term, and
$\phi$ is the vev of $\phi_1$ multiplied by $\phi_1 \psi
\bar\psi_V$ coupling. This is simply because the low energy third
generation field is a linear combination of the form $\cos
\alpha\, \psi_3- \sin \alpha\, \psi_V$  with the mixing angle
$\sin \alpha \simeq \phi/\sqrt{M_1^2+\phi^2}$.

 Therefore, in general,
there is no gross contradiction to the fact that the top Yukawa
coupling is order 1. However, in our case, if $\phi/M_1$ becomes
close to 1, the atmospheric mixing shifts from the maximal angle.
Therefore, that needs to be addressed if the precise
tri-bi-maximal mixing and $h_{\psi_V\psi_V H} \alt 1$ is demanded.
%
%
 The desired smallness of the effective $f$ and $h^\prime$ couplings
however are more naturally obtained due to the presence of the
Planck mass in the denominator. In order to make the $f$-coupling
dominate over the $h^\prime$, we have to choose a small coupling
for the $H^\prime$ Higgs field in Eq. (4). Similarly the $\lambda$
term in Eq. (9) is assumed to be small compared to the coefficient
of the first matrix.

Thus within these set of assumptions, this model is in good
phenomenological agreement with observations. In a more complete
theory, these assumptions need to be addressed. We however find it
remarkable that despite these shortcomings, the model provides a
very useful unification strategy of the diverse quark-lepton
mixing patterns.

 \section{Vacuum alignment}
 The major new point of this note is that we obtain the above
 fermion mass matrices from an $S_4$
 symmetry where the minimum configuration of the
 flavon fields used in our analyse of fermion mixings
  arise from superpotential minimization with very additional
  assumptions.

We start our discussion by giving some simple examples and
discussing the flavon alignment as a prelude to the more realistic
example. First thing to note is that ${\bf 3_1}^3$ is invariant
under $S_4$, but ${\bf 3_2}^3$ is not. Denoting $\phi = (x,y,z)$,
we see that in the first case, the singlet of $\phi^3 = xyz$. The
superpotential for a $\bf 3_1$ flavon field $\phi$ can therefore
be written as
\begin{equation}
W= \frac12 m \phi^2 - \lambda \phi^3 = \frac12m(x^2+y^2+z^2) -
\lambda xyz.
\end{equation}
The solution of $F$-flat vacua ($\phi\neq 0$) are
\begin{equation}
\phi = \frac{m}{\lambda} \{(1,1,1)\ {\rm or}\ (1,-1,-1)\ {\rm or}\
(-1,1,-1)\ {\rm or}\ (-1,-1,1)\}.
\end{equation}
These aligned vacua can be identified to the vertex diagonal axes of
the regular hexahedron.
 In fact $S_4$ can be
identified the permutation of the 4 axes of regular hexahedron.
Once one of the axes is fixed, $S_3$-permutation is left.
Therefore, the vacua break $S_4$ down to $S_3$.


On the other hand,
when $\bf 3_2$ flavon is used (or the cubic term
is forbidden by a discrete symmetry),
quartic term involving the triplet is crucial for the $F$-flat vacua.
The invariant quartic term $\phi^4$
gives two linear combinations of the form
$x^4+y^4+z^4$ and $x^2 y^2+ y^2 z^2+z^2 x^2$.
This is because they have to be symmetric homogenous terms and invariant
under the Klein's group, which is $\pi$ rotation around the $x,y,z$ axes.
%
%
%
%
%
%
Thus, the superpotential
term for $\bf 3_2$ field $\phi$ is
\begin{eqnarray}
W &=& \frac12 m \phi^2 - \frac{\kappa^{(1)}}{M} (\phi^4)_1 -
\frac{\kappa^{(2)}}{M} (\phi^4)_2 \\
&=& \frac12(x^2+y^2+z^2)- \frac{\kappa^{(1)}}{4M} (x^4+y^4+z^4) -
\frac{\kappa^{(2)}}{2M}(x^2y^2+y^2z^2+z^2x^2). \nonumber
\end{eqnarray}
The nontrivial $F$-flat vacua ($\phi\neq 0$) are
\begin{equation}
\phi = \sqrt{\frac{mM}{\kappa^{(1)}}}\, \vec{a},\quad
       \sqrt{\frac{mM}{\kappa^{(1)}+2\kappa^{(2)}}}\, \vec{b}, \quad
       \sqrt{\frac{mM}{\kappa^{(1)}+\kappa^{(2)}}}\, \vec{c},
\label{the-vacua}
\end{equation}
where $\vec{a} = (0,0,\pm1)$, $(0,\pm1,0)$, $(\pm1,0,0)$, $\vec{b}
= (\pm1,\pm1,\pm1)$, and $\vec{c} = (0,\pm1,\pm1)$,
$(\pm1,\pm1,0)$, $(\pm1,0,\pm1)$. We note that these vectors
correspond to the axes of the regular hexahedron. The vacua break
$S_4$ down to $Z_4$, $Z_3$, and $Z_2$, respectively. More
importantly, the vacuum states in Eq. (7) used in the analysis of
fermion masses in the previous section are a subset of the above
vacua.
%

Note that if we add a $\phi^4$ term to the superpotential involving the $\bf 3_1$ flavon field,
$\vec{a}$ vacuum is possible, in addition to the original $\vec{b}$ vacua.
However, $\vec{c}$ vacuum is absent. 

Turning to the model at hand, due to non-trivial $Z_n$ 
charges for the flavon fields, the mass terms (bilinears) are not allowed. To solve this problem,
 we have included $\bar\phi$ fields which then lead to Dirac type mass terms.
The superpotential for a single flavon field is then given
$W_i= m_i \phi_i\bar \phi_i + \kappa_i \phi_i^2\bar\phi_i^2$. 
Denoting $\phi_i=(x_i,y_i,z_i)$ and $\bar\phi_i=(\bar x_i,\bar y_i,\bar z_i)$. In terms of its component fields,
we obtain
\begin{eqnarray}
W_i &=& m_i (x_i\bar x_i + y_i \bar y_i+ z_i \bar z_i)
+ \frac{\kappa_i^{(1)}}{M} (x_i^2 \bar x_i^2 + y_i^2 \bar y_i^2 + z_i^2 \bar z_i^2) \\
&&+ \frac{\kappa_i^{(2)}}{M} \left(x_i^2 (\bar y_i^2 + \bar z_i^2) +
y_i^2 (\bar z_i^2+\bar x_i^2) + z_i^2 (\bar x_i^2+\bar y_i^2)\right)
+ \frac{\kappa_i^{(3)}}{M} (x_i\bar x_i y_i \bar y_i+ y_i \bar y_i z_i \bar z_i+
z_i\bar z_i x_i \bar x_i). \nonumber
\end{eqnarray}
Note that there are three kinds of invariant for $\phi^2 \bar\phi^2$.
%
Finding the $F$-flat solution of this superpotential
is similar to the case in Eq.(\ref{the-vacua}).
It is easily verified 
that the $F$-flat vacua are proportional to $\vec{a}$, $\vec{b}$, and $\vec{c}$
similarly in Eq.(\ref{the-vacua}).

Several comments are now in order:

\begin{itemize}

\item We note that the cubic terms in the flavon superpotential,
such as $\phi_2^3$, $\phi_2^2 \bar \phi_3$ are forbidden by our
choice of $Z_n$ charge of $s_i$ since their presence will spoil a
vacuum alignment of $\phi_2$ in Eq.(\ref{flavon}).
\item Secondly, note that the orthogonality of the vevs of
$\phi_2$ and $\phi_3$
 is important to obtain the tri-bi-maximal mixing . One way to
obtain it dynamically is to have
a mixing term $\phi_2^2 \phi_3^2$ such that the coupling of the
mixing term is much smaller than $\phi_2^2\bar\phi_2^2$ and
$\phi_3^2\bar\phi_3^2$ couplings. The invariant term
$\phi_2^2\phi_3^2$ expressed in terms of components gives $x_2 x_3
y_2 y_3 + y_2 y_3 z_2 z_3 + z_2 z_3 x_2 x_3$, where $\phi_2 =
(x_2,y_2,z_2)$ and $\phi_3 = (x_3,y_3,z_3)$. The $F$-flatness
condition implies that $y_2 y_3 + z_2 z_3 = 0$ when $x_2=0$ and
$x_3 \neq 0$ leading to the desired orthogonality of the
alignments of $\langle \phi_2\rangle$ and $\langle\phi_3\rangle$.
Note that with our $Z_n$ charge assignments, this can arise only
in higher orders and its coefficients must therefore be small. The
same situatrion happens also for the mixing terms of the form:
$\phi_2\bar\phi_2\phi_3\bar\phi_3$\footnote{ An alternative way to
obtain the above orthogonality is to make a different choice of
reps of $S_4$ i.e. introduce an extra flavon field $\phi_4$
 whose vev is $\phi_4^t =(1,0,0)$. The rank one $f$ coupling is obtained
by $(\phi_3 \phi_4 \psi) \bar \psi_V + M_V \psi_V \bar \psi_V +
\psi_V \psi_V \bar\Delta$ instead of $\phi_2\bar \psi_V + M_V
\psi_V \bar \psi_V + \psi_V \psi_V \bar\Delta$. Suppose under
$S_4$ $\phi_3$, $\phi_4$ and $\psi$ transform as $\bf 3_1$, $\bf
3_1$, and $\bf 3_2$, respectively. Then, if the reps of $\psi_V$,
$\bar\psi_V$ are $\bf 1_1$, the product of $\phi_3$ and $\phi_4$
should be $\bf 3_2$, which is obtained by anti-symmetric product.
At that time, the orthogonal condition is satisfied automatically
 (i.e., $\phi_3 \times \phi_4 \perp \phi_3$).}.

 \item There are mixing terms between the different flavon
fields in the quartic terms of the form
$W_{ij}~=~\frac{\lambda}{M}\phi_i\bar\phi_i \phi_j\bar\phi_j$.
When expressed in terms of the component fields $x,y,z$, they
involve mixed terms like $\lambda(x_i\bar x_i y_j \bar y_j+ y_i
\bar y_i z_j \bar z_j+ z_i\bar z_i x_j \bar x_j)$ plus similar
other mixed invariants. In the previous item, we just discussed
the case when $i=2$ and $j=3$. As for the remaining terms of this
type, they will in general induce small contributions proportional
to $\lambda$ in the vevs in Eq.(7) where there are zeros. They
will induce correction to the forms of our mass matrices.  We will
therefore need to assume that these $\lambda$ couplings to be
small, so that their effect on our mass and mixing predictions
will be small.

\item Depending on the values of $a$ and $b$, one could in
principle get very high dimensional terms of the form $s_1^x s_2^y
\phi_2\phi_2 \bar{\phi}_3$ ($x,y$ are positive integers); however
their contribution to the flavon potential is suppressed and we
ignore these effects.

\end{itemize}

We therefore conclude that all the desired vacua  in the SO(10)
model are present. Any possible corrections to them can be made
small making it possible to take a first step towards building a
unified model of flavor.


%

%

\section{Conclusion}
In summary, we have proposed a grand unified model for
quark-lepton flavor starting above the GUT scale with an SO(10)
theory with $S_4\times Z_n$ 
discrete symmetry, $S_4$ non-singlet flavon fields and two vector
like pairs of {\bf 16} with mass above the GUT scale and SO(10)
Higgs multiplets {\bf 10} and {\bf 126} fields that give mass to
fermions. The {\bf 16} matter as well as the flavon fields
transform as $S_4$-family group triplets. The ground state of the
flavon sector of the theory gives non-zero vevs to the flavon
fields along specific directions due to the above discrete
symmetries and when certain higher dimensional couplings between
different flavon fields are assumed to be small. They fix the
structure of the Yukawa couplings of {\bf 10} and {\bf 126} fields
at GUT scale after the vector-like fields decouple. This leads to
specific mass textures for the quarks and leptons with only a few
parameters and hence the predictions for quark lepton mass
relations and mixing angles in both the quark and the lepton
sector. In particular, the model leads to tri-bi-maximal form for
the PMNS matrix in the leading order with corrections to this
coming from charged lepton fields. Using this, we predict
$\theta_{13}\simeq 0.05$. The quark mass hierarchies as well as
quark mixings given by the model are in agreement with
observations e.g. the model predicts at GUT scale correct mass
ratios for $m_b/m_\tau$ and $m_s/m_\mu$ as well as the Cabibbo
angle $V_{us}$ without any adjustment of parameters. Some
assumptions are needed to get the large top quark Yukawa coupling
as well as relative strengths between the various flavon
couplings. Clearly, our work begins a process which seems very
promising and further work is needed to improve some of the
assumptions used.

\appendix
\section*{Appendix : $S_4$ group}

We briefly review the $S_4$ group.
The group $S_4 \simeq D_2 \rtimes D_3 \simeq (Z_2 \times Z_2)
\rtimes S_3$ has irreducible reps $\bf 1_1$, $\bf 1_2$, $\bf
2$, $\bf 3_1$ and $\bf 3_2$ as noted. To see the detailed
properties, we use the $(x,y,z)$ coordinate for the transformation law of the
three-dimensional representations of $S_4$. The group $Z_2 \times Z_2$ is a
Klein's group $K = \{ {\rm diag} (1,1,1), {\rm diag}(1,-1,-1),
{\rm diag}(-1,1,-1), {\rm diag}(-1,-1,1)\}$, which corresponds to
$\pi$ rotation around the $x,y,z$ axes. The group $S_3$ is a permutations of
the three axes $(x,y,z)$: $S = \{ {\rm diag} (1,1,1),
\left(\begin{array}{ccc} 0 &1 & 0 \\ 0 &0 & 1\\ 1 & 0 & 0
\end{array} \right) \left(\begin{array}{ccc} 0 &0 & 1 \\ 1 &0 &
0\\ 0 & 1 & 0 \end{array} \right) \left(\begin{array}{ccc} 1 &0 &
0 \\ 0 &0 & 1\\ 0 & 1 & 0 \end{array} \right)
\left(\begin{array}{ccc} 0 &0 & 1 \\ 0 &1 & 0\\ 1 & 0 & 0
\end{array} \right) \left(\begin{array}{ccc} 0 &1 & 0 \\ 1 &0 &
0\\ 0 & 0 & 1 \end{array} \right) \}$. The element of $S_4$ is
given as $S_4 = \{ (k,s)| k\in K, s \in S \}$.

The $\bf 3_1$
representation $\phi$ (column vector) transforms by the action of
$S_4$ as
\begin{equation}
\phi \to ks \phi,
\end{equation}
while $\bf 3_2$ representation $\phi^\prime$ transforms as
\begin{equation}
\phi^\prime \to (\det s) ks \phi^\prime.
\end{equation}
The singlet $\bf 1_2$ transforms as
\begin{equation}
{\bf 1_2} \to (\det s) {\bf 1_2},
\end{equation}
and $\bf 1_1$ is invariant under the action of $S_4$.
The reps $\bf 1_1$, $\bf 1_2$, $\bf 2$ are reps of $S_3 \simeq D_3$, and the
transformation law of $\bf 2$ is rotation and reflection
of the regular triangle.
Doublet rep $(u,v)$ transforms as
\begin{equation}
\left(
\begin{array}{c}
u \\
1 \\
v
\end{array}
\right) \to U^t_{\rm TB}\, s\, U_{\rm TB} \left(
\begin{array}{c}
u \\
1 \\
v
\end{array}
\right).
\end{equation}

For convenience, we list the Kronecker products of the triplets:
\begin{eqnarray}
&&({\bf 3_i \times 3_i})_s = {\bf 1_1} \oplus {\bf 2} \oplus {\bf 3_1}, \quad
({\bf 3_i \times 3_i})_a = {\bf 3_2}, \nonumber \\
&&{\bf 3_1 \times 3_2} = {\bf 1_2} \oplus {\bf 2} \oplus {\bf 3_1} \oplus {\bf 3_2}. \nonumber
\end{eqnarray}

\section*{Acknowledgement}
The work of R.~N.~M. and Y. M. is supported by the US National
Science Foundation under grant No. PHY-0652363  and that of B. D.
is supported in part by the DOE grant DE-FG02-95ER40917. Y. M.
acknowledges partial support from the Maryland Center for Fundamental
Physics. We thank G. Altarelli for discussions.

\end{document}